\documentclass{aastex631}
\usepackage{bm}

\begin{document}

\title{The Evolution of the 1/f Range Within a Single Fast-Solar-Wind Stream Between 17.4 and 45.7 Solar Radii
}f


\author[0000-0001-7222-3869]{Nooshin Davis}
\author[0000-0003-4177-3328]{B. D. G. Chandran}
\affiliation{
Department of Physics and Astronomy, University of New Hampshire, Durham, New Hampshire 03824, USA}

\author[0000-0002-4625-3332]{T. A. Bowen}
\affiliation{Space Sciences Laboratory, University of California, Berkeley, CA 94720-7450, USA}
\author[0000-0002-6145-436X]{S. T. Badman}
\affiliation{Center for Astrophysics, Harvard and Smithsonian, Cambridge, MA, USA}
\author[0000-0002-4401-0943]{T. Dudok de Wit}
\affiliation{LPC2E, CNRS and University of Orléans, Orléans, France}
\affiliation{International Space Science Institute, ISSI, Bern, Switzerland }
\author[0000-0003-4529-3620]{C. H. K. Chen}
\affiliation{Department of Physics and Astronomy, Queen Mary University of London, London E1 4NS, UK}
\author[0000-0002-1989-3596]{ S. D. Bale}
\author[0000-0001-9570-5975]{Zesen Huang }
\author[0000-0002-1128-9685]{Nikos Sioulas}
\author[0000-0002-2381-3106]{Marco Velli}
\affiliation{Department of Earth, Planetary, and Space Sciences, University of California, Los Angeles, CA, USA}

\begin{abstract}

The  power spectrum of magnetic-field fluctuations in the fast solar wind ($V_{\rm SW}> 500 \mbox{ km} \mbox{ s}^{-1}$) at magnetohydrodynamic (MHD) scales is characterized by two different power laws on either side of a break frequency~$f_{\rm b}$. The low-frequency range at frequencies~$f$ smaller than $f_{\rm b}$ is often viewed as the energy reservoir that feeds the turbulent cascade at $f>f_{\rm b}$. At heliocentric distances~$r$ exceeding 60 solar radii~($R_{\rm s}$), the power spectrum often has a $1/f$ scaling at $f<f_{\rm b}$; i.e., the spectral index is close to~$-1$.
In this study, measurements from the {\em Parker Solar Probe}’s (PSP's) encounter 10 with the Sun are used to investigate the evolution of the magnetic-field power spectrum at $f< f_{\rm b}$ at $r<60 R_{\rm s}$  during a fast radial scan of a single fast-solar-wind stream. We find that the spectral index in the low-frequency part of the spectrum decreases from approximately -0.61 to -0.94 as $r$ increases from $17.4 $ to $45.7$ solar radii. 
Our results suggest that the $1/f $ spectrum that is often seen at large~$r$ in the fast solar wind is not produced at the Sun, but instead develops dynamically as the wind expands outward from the corona into the interplanetary medium.

\end{abstract}




\section{Introduction} \label{sec:intro}

The heliosphere is permeated by the solar wind, a supersonic and super-Alfvenic plasma flow of solar origin that continually expands into the heliosphere. Throughout its radial expansion, the solar wind has a strongly turbulent character \citep{Bruno13}, with spatial and temporal variations over a wide range of scales \citep[see, e.g.,][]{goldstein95a,Verscharen19}. For example, 
at frequencies~$f$ in the spacecraft frame that correspond (via's Taylor's (1938)  \nocite{taylor38} hypothesis) to ``magnetohydrodynamic (MHD)'' length scales (i.e., length scales much larger than the proton gyroradius),
the power spectral density (PSD) observed in the fast solar wind near Earth displays different power-law scalings $f^{\alpha_B}$ on either side of a break frequency~$f_{\rm b} \sim 10^{-3} \mbox{ s}^{-1}$ \citep{Bruno13}. At $f< f_{\rm b}$,  $\alpha_B$ is often $\simeq -1$,  especially in the fast solar wind  \citep{tumarsch95,Bruno13}. 
At $f > f_{\rm b}$,  $\alpha_B$ is approximately $-5/3$ to $-3/2$. Such a spectral index, combined with
the observed anisotropy of solar-wind turbulence~\citep[e.g.][]{matthaeus90,horbury08,sahraoui10,wicks11,chen12}, is consistent with theories of anisotropic MHD turbulence, including the possible presence of dynamic alignment and intermittency \citep{goldreich95, maron01, boldyrev05, beresnyak08, chandran15, mallet17a,Schekochihin22}.

The range $f< f_{\rm b}$ is sometimes called  the ``energy-containing range,'' and the corresponding length scales are sometimes called the ``energy injection scales.''
The origin of the $1/f $ scaling at $f< f_{\rm b}$, which is often observed in fast solar wind not only near Earth but also more generally at $r > 60 R_{\rm s}$,
is still under debate \citep{Bruno19}. \cite{roberts89} found that the amplitudes of the fluctuations at $f<f_{\rm b}$ evolve between 0.3~au and 1~au in the same way as Alfv\'en waves undergoing WKB propagation without turbulent decay. He then suggested that if no nonlinear decay occurs between 0.3~au and 1~au, then no decay should occur between the corona and 0.3~au, as the fractional variations in $B$ are smaller closer to the Sun. If correct, his arguments would imply that the $1/f$ part of the spectrum is produced at the Sun. \cite{matthaeus86} offered a different argument that the $1/f$ scaling is produced at the Sun, suggesting that the $1/f$ scaling results from superposing different, uncorrelated samples of fluctuations originating from different regions of the solar surface. Other studies have taken an opposing point of view, attributing the $1/f$ scaling in the fast solar wind to turbulent dynamics within the solar wind. For example, some studies have linked the $1/f$ spectrum to reflection-driven MHD turbulence \citep{velli89,verdini12}, in which non-WKB reflection is the primary source of the Sunward-propagating Alfv\'en waves that interact nonlinearly with the dominant outward-propagating Alfv\'en waves. 
It has also been suggested that the $1/f$ scaling seen by the {\em Helios} spacecraft in the fast solar wind at $f \ge 3 \times 10^{-4} \mbox{ s}^{-1}$ \citep{tumarsch95} is produced in situ during the nonlinear evolution of the  parametric decay instability \citep{chandran18a}.\cite{matteini_1_2018} 
proposed that the existence of a 1/f spectrum in Alfv\'enic fast streams is  associated with the presence of an observational cutoff in the distribution of the fluctuations and the saturation of their mean amplitude. 

In this paper, we use measurements from the {\em Parker Solar Probe} (PSP) to investigate the evolution of the magnetic power spectral density at $f< f_{\rm b}$ as  $r$ increases from $17 R_{\rm s}$ to $45 R_{\rm s}$. We focus on a data interval during which PSP executed a fast radial scan of a single fast-solar wind stream, so that the radial variations of solar-wind properties during this interval can be plausibly mapped to temporal variations of those properties in the solar-wind rest frame. Section~\ref{sec:data} describes the data we analyze in more detail, as well as the magnetic connectivity of PSP to the Sun during the period we consider. In Section~\ref{sec:result}, we present our main results, and in  Section~\ref{sec:summary} we summarize our findings and conclude.

\section{ Data } \label{sec:data}

In this study, we analyze magnetic-field and velocity data from PSP Encounter 10 from Nov. 17-20, 2021, covering a heliocentric radial distance range 0.08–0.22 au (17.4 to  45.7 $R_s$). 
Over the entirety of PSP's encounter 10, PSP connected to a series of three isolated mid-latitude negative-polarity coronal holes \citep{badman23}. PSP traversed all three in rapid succession as it moved prograde over the solar surface near perihelion. Most of this motion, however, occurred over the course of 2 days from November 21 - November 23 2021, during which PSP moved nearly 90 degrees in longitude. Prior to this, and most relevant for the present study, PSP was near corotation with the Sun while moving inwards rapidly: from November 17-November 20, PSP's heliographic location varied by less than 10 degrees in Carrington longitude, and field-line mapping places the source region near the center of the same coronal hole for this entire period. From November 17 to November 19, the footpoints moved very slighly in retrograde, away from the center (peak) of the fast wind stream and towards the East limb. Over the remainder of November 19, PSP went through corotation and then began to migrate slowly prograde before passing over the fastest part of the stream around midday on November 20th \citep{badman23}. The data interval that we analyze basically provides a radial scan of the same fast-solar-wind stream. Note that this interval excludes not only crossings of the heliospheric current sheet, but also coronal mass ejections.

For this study, we used magnetic-field data from the outboard fluxgate magnetometer (MAG) from the FIELDS instrument suite (Bale et al. 2016) at a resolution of 0.21845~s. The velocity data are derived from Solar Wind Electrons Alphas and Protons (SWEAP) \citep{kasper16} measurements made by the Solar Probe Analyzers (SPAN) at a resolution of 3.491~s \citep{Livi22}.

\begin{figure}[ht!]
\centering

\includegraphics[width=5in,height=5in]{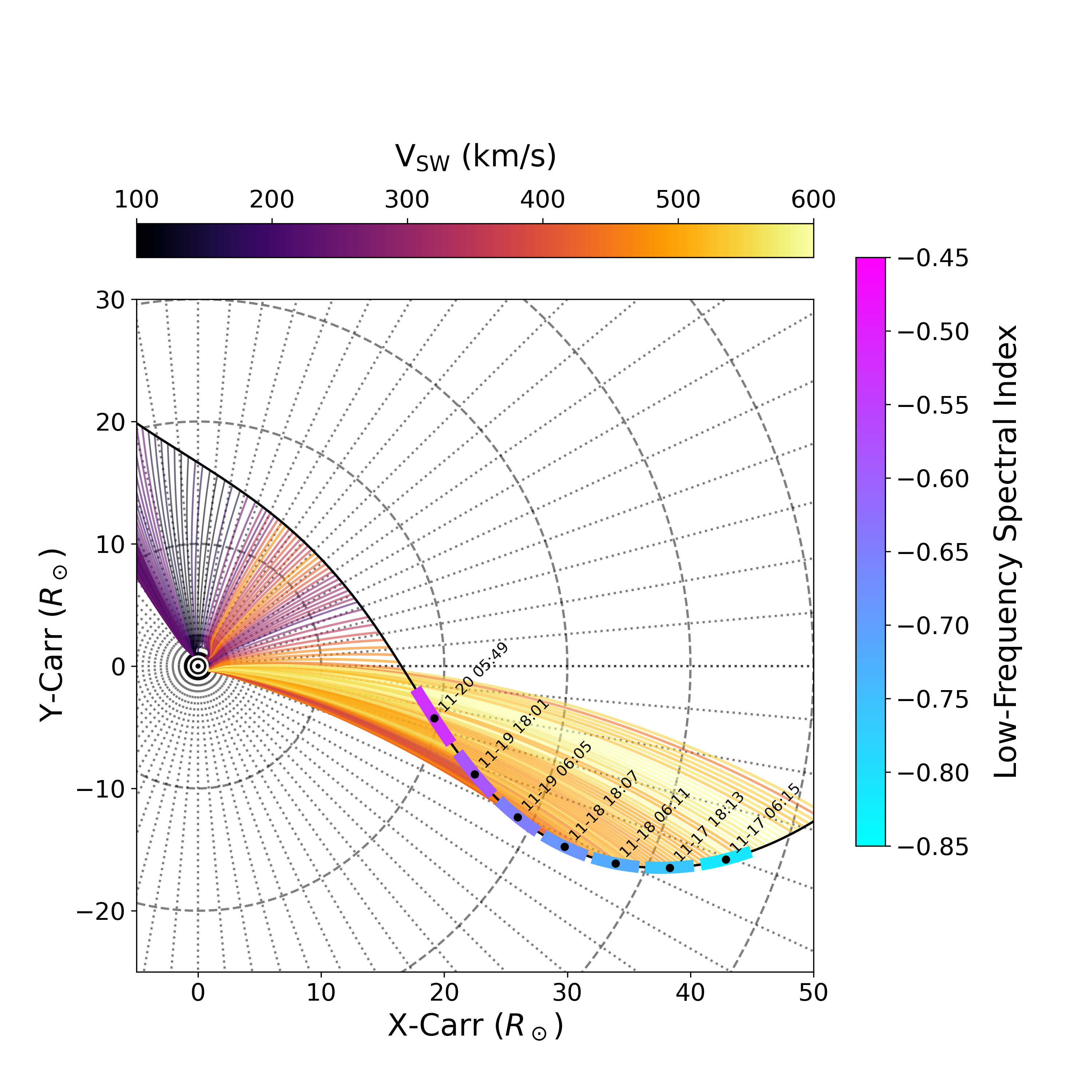}
\caption{Encounter 10 orbital trajectory of Parker Solar Probe from Nov.17 to mid Nov.20 plotted in Carrington (solar-corotating) coordinates. Colored blocks along the trajectory are color coded by magnetic spectral index. Parker spiral magnetic field lines colored by the measured radial solar wind velocity show that the portion of the trajectory studied in this work remains in the same fast solar wind stream while traveling nearly radially with respect to the Sun.}
\label{fig:fig1}
\end{figure}

\begin{figure}[ht!]
\centering
\includegraphics[width=6in,height=3.5in]{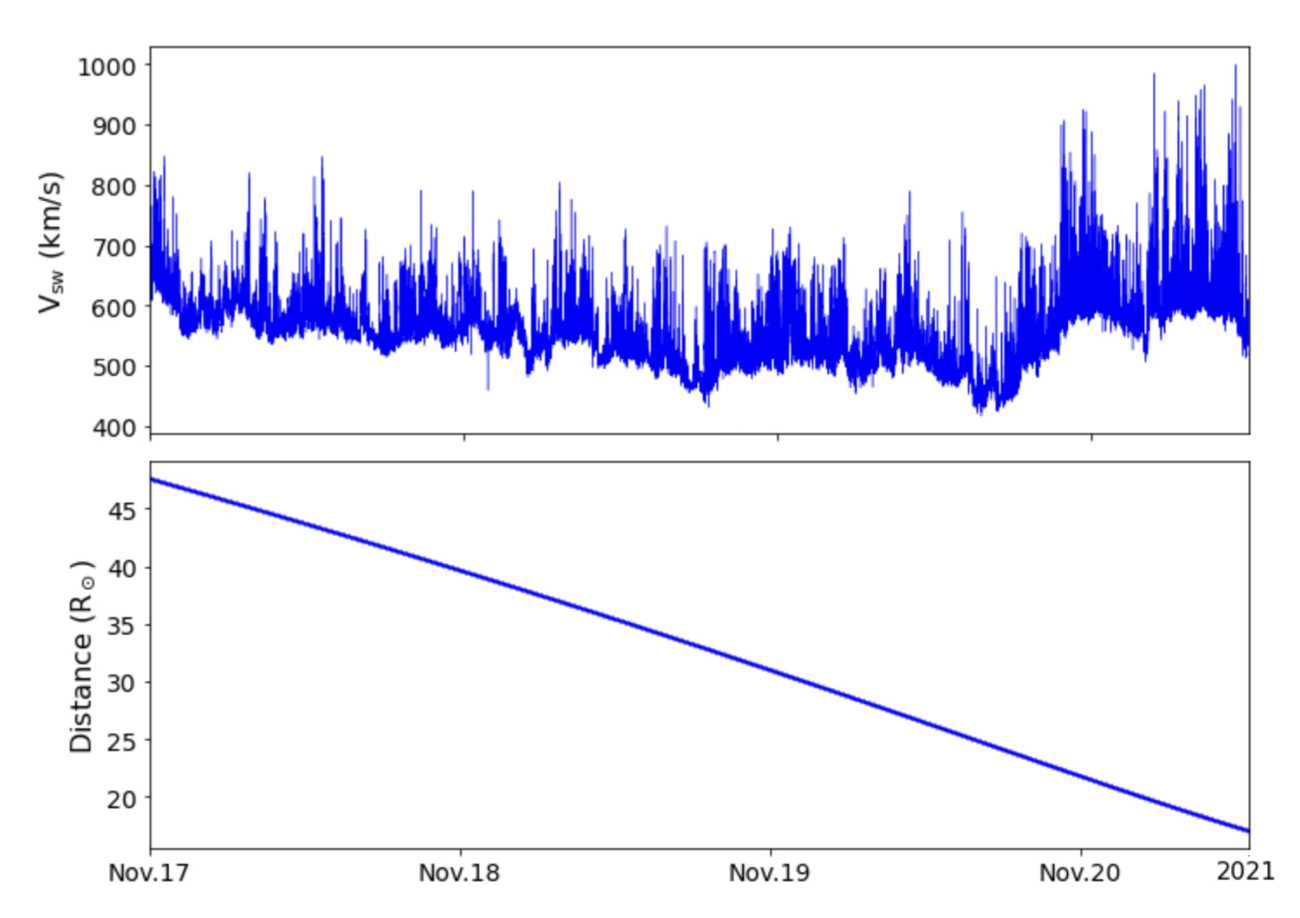}
\caption{(Top) Variation of the radial solar wind velocity (Bottom) Heliocentric distance of Parker Solar Probe. }
\label{fig:fig2}
\end{figure}

 Figure~\ref{fig:fig1} illustrates the trajectory of the probe over the period we analyze. As this figure shows, during this time interval, PSP was traveling approximately radially from 45.7 to 17.4 $R_s$. Also shown in  Figure~\ref{fig:fig1} is the radial velocity of the solar wind, denoted by $V_{\rm sw}$, with lighter colors corresponding to the fast solar wind. All radial velocities that we show in this paper are in the Sun's rest frame, not the spacecraft frame. Figure \ref{fig:fig2} reproduces the data for $V_{\rm sw}$ in a different format, along with the heliocentric distance of PSP, both as functions of time. As shown in both figures, over this time interval, the solar wind was fast solar wind, with $V_{\rm sw}$ exceeding $500 \mbox{ km} \mbox{ s}^{-1}$ the grand majority of the period.
 
\section{Results} \label{sec:result}
\begin{figure}[ht!]
\centering
\includegraphics[scale=0.6]{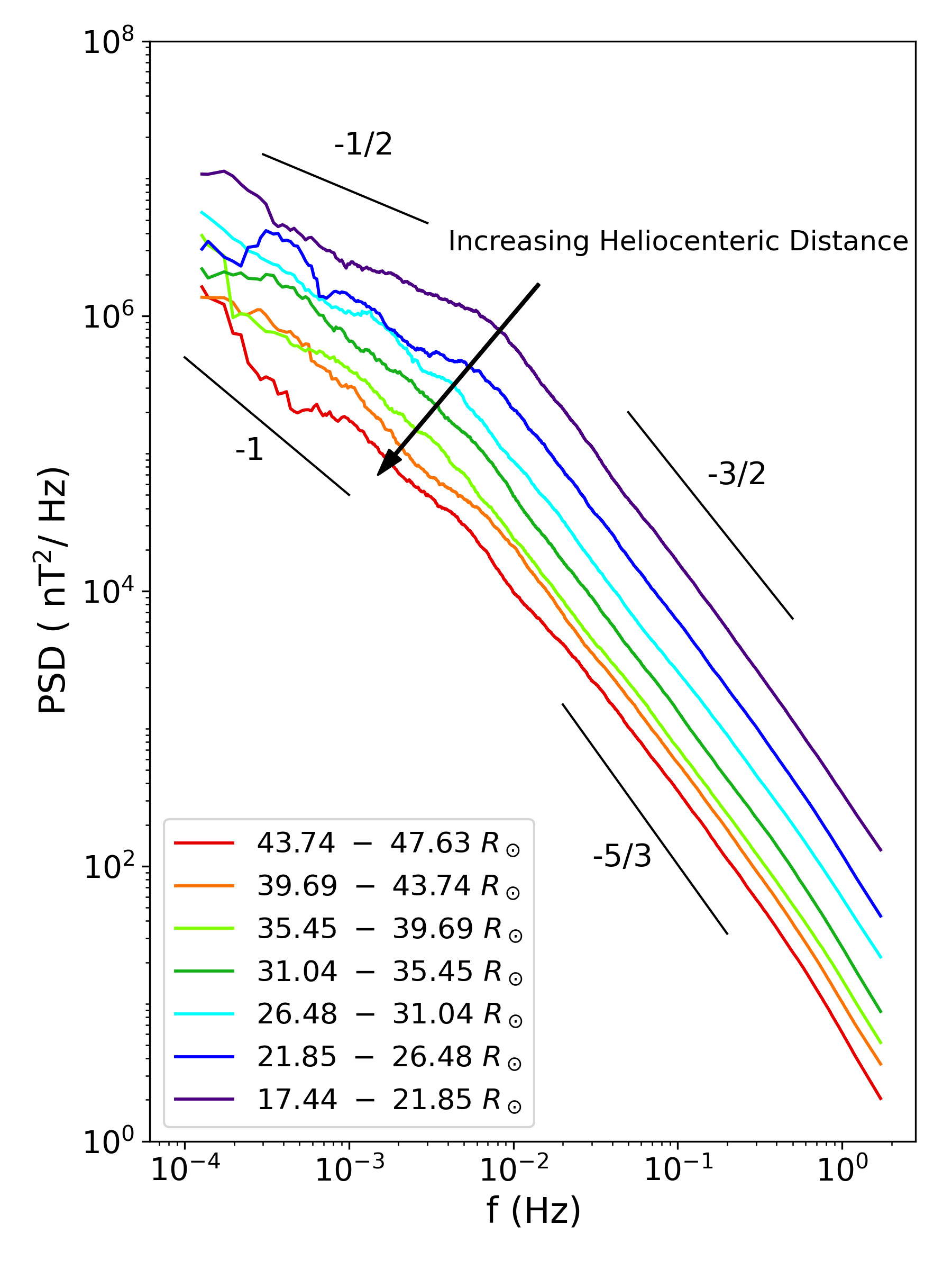}
\caption{Magnetic field power spectrum for different heliocentric distances. Several power-law slopes is marked for comparison. } 
\label{fig:fig3}
\end{figure}

One common method to measure the multi-scale nature of turbulence is via the Power Spectral Density (PSD) of the turbulent fluctuations as a function of the spacecraft-frame frequency. To examine the evolution of the magnetic-field fluctuation spectrum and the low-frequency spectral properties of solar wind observed by PSP, we divide the MAG data into 12-hour intervals.
For each interval, we employ a Fourier transform to build the PSD of the magnetic fluctuations. 
To improve the clarity of our plots, we average the spectra over a sliding window of a factor of 2 in the frequency domain. The power spectra for the different 12-hour intervals are shown in Figure~\ref{fig:fig3}, with each spectrum colored by heliocentric distance. It can be seen that the power levels systematically decrease with increasing heliocentric distance, with a total decrease exceeding one order of magnitude over the range of distances considered. This behavior is consistent with observations at~$r> 0.3 \mbox{ au}$ and is due to the expansion of the solar wind and the turbulent cascade \citep{heinemann80,tumarsch95}.

Throughout the range of distances studied, the power spectra show a power-law range compatible with models of inertial-range MHD turbulence for frequencies exceeding the break frequency~$f_{\rm b}$ introduced in Section~\ref{sec:intro}, which is~$\sim 10^{-2} \mbox{ s}^{-1}$ for the top few spectra shown in Figure~\ref{fig:fig3}. The spectral indices at $f> f_{\rm b}$ fall within the expected range $(-5/3, -3/2)$ predicted in theories of MHD turbulence \citep[e.g.][]{goldreich95,maron01,boldyrev05,beresnyak08,chandran15,mallet17a} and recent PSP observations \citep{chen20,sioulas23}, but this frequency range is not the focus of the present study. Figure \ref{fig:fig3} marks several power-law slopes for comparison.

\begin{figure}[ht!]
 \centering
 \plotone{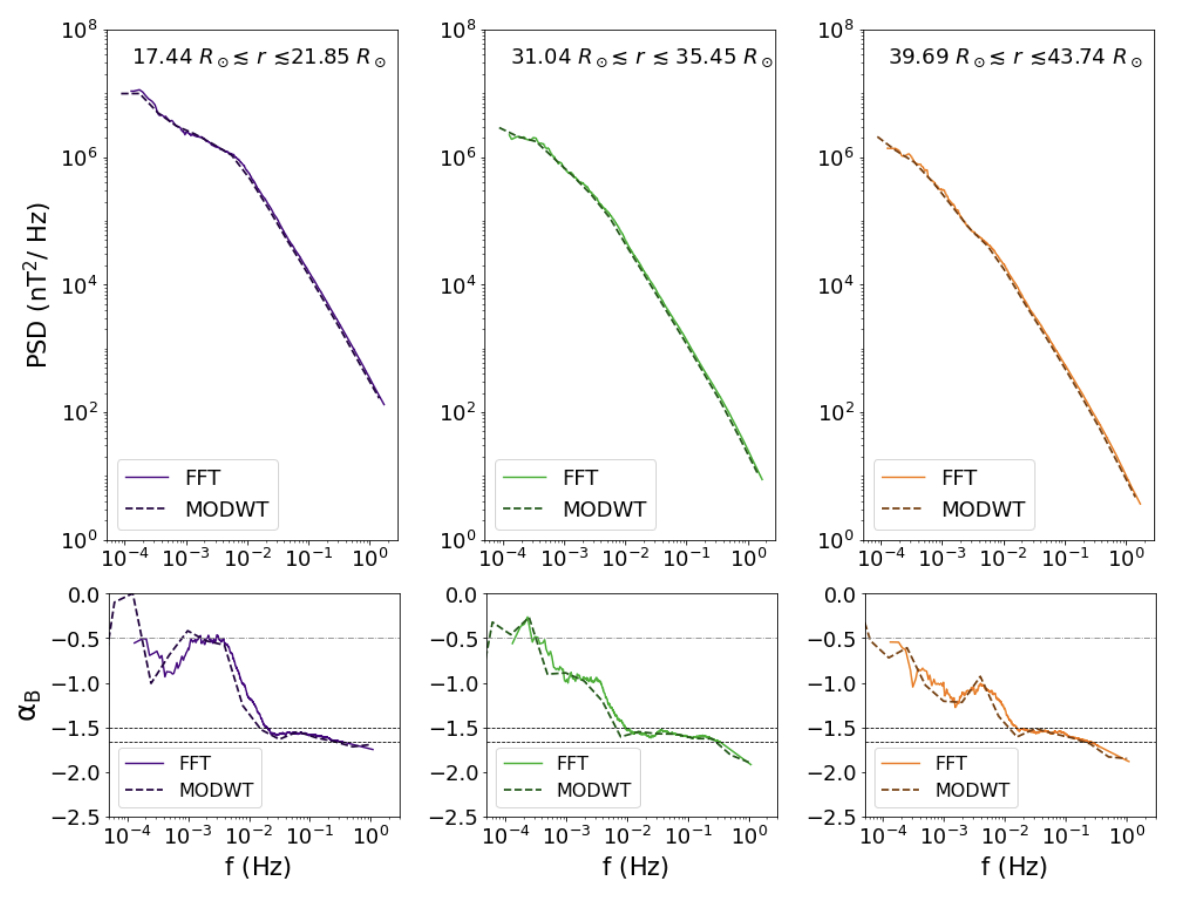}

\caption{(Top) Magnetic field power spectra and
(Bottom) Local spectral index, for different heliocentric distances. The horizontal dotted lines mark the values $\alpha_{\rm B} = -1/2$, $-3/2$, and~$-5/3$.}
\label{fig:fig4}
\end{figure}

Figure \ref{fig:fig4} shows the power spectra for three different heliocentric distance ranges.
This figure also shows the local spectral index~$\alpha_B(f)$, which we compute by determining the best linear fit to the power spectrum in $\log(\mbox{PSD}) - \log(f)$ space in the frequency interval $(f/ \sqrt{10}, \sqrt{10} f)$ -- i.e., over a factor of 10 in frequency. 
As $r$ increases from $\simeq 20 R_{\rm s}$ to $\simeq 40 R_{\rm s}$, the value of $\alpha_B$ at $f<f_{\rm b}$ decreases from $\simeq -0.5$ to $\simeq -1$;  that is, the power spectrum at low frequency gradually steepens towards a $1/f$ spectrum. It can also be seen in Figure~\ref{fig:fig4} that $f_{\rm b}$ decreases as~$r$ increases, as also happens at larger~$r$ \citep{Bruno13,chen20}. Note that the low frequency power laws are not perfect power laws, possibly due to the limited statistics, but that the difference in the slopes can still clearly be seen.

\begin{figure}[ht!]
\centering
\plotone{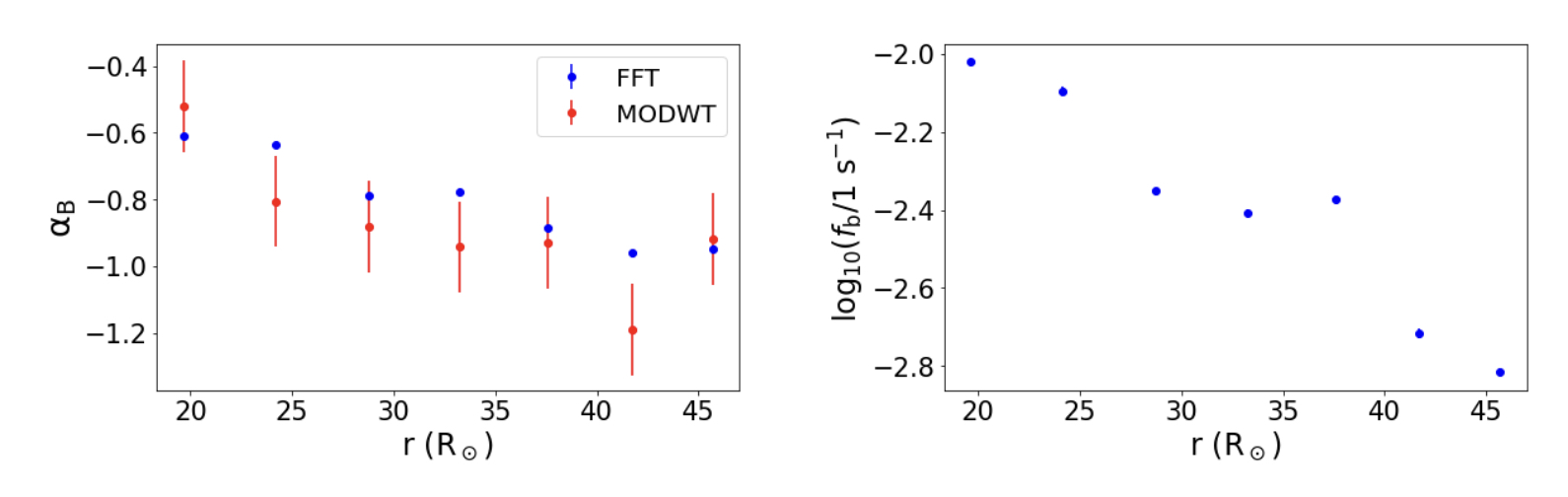}
\caption{(Left) Variation of the magnetic field spectral index and (Right) spectral break point, with heliocentric distance. Blue points refer to the FFT approach and orange triangles refer to the MODWT approach.}
\label{fig:fig5}
\end{figure}

Figure~\ref{fig:fig5} illustrates the radial evolution of $f_{\rm b}$ and the value of $\alpha_B$ at $f< f_{\rm b}$ using all seven 12-hour intervals.
For this plot, we employ a method to compute $\alpha_B$ that differs from the one used in Figure~\ref{fig:fig4}. In particular, we fit the power spectra to a double power-law and adjust the following four parameters simultaneously to optimize the fit: the break frequency~$f_{\rm b}$, the amplitude of the power spectrum at $f=f_{\rm b}$, the spectral index at $f<f_{\rm b}$, and the spectral index at $f>f_{\rm b}$. The spectral index at $f<f_{\rm b}$ from this procedure is denoted by $\alpha_B$ in Figure~\ref{fig:fig5}. As shown in this figure, the spectral index at $f<f_{\rm b}$
decreases from -0.61 at 17.4~$R_s$ to -0.94 at 45.7~$R_s$. Over the same radial interval, the break point frequency decreases from $7.4 \times 10^{-3}$ to $2.7 \times 10^{-3}$ Hz. Our results suggest that the energy-containing range of the spectrum develops dynamically within the fast solar wind over this range of radii, and that the $1/f$ spectrum seen in fast solar wind at $r>0.3 \mbox{ au}$ is not produced at the Sun.

To explore the extent to which our results may depend on our analysis technique, we re-analyze the data using a wavelet transform. 
Specifically, we use the maximal overlap discrete wavelet transform (MODWT) \citep{Percival12}, which provides direct estimates of the spectral index and its confidence interval for individual octaves (factors of 2 in frequency). 
We then determine an effective spectral index at $f<f_{\rm b}$ by averaging the wavelet spectral indices over the frequency range $(4\times 10^{-4} \mbox{ s}^{-1}, 3 \times 10^{3} \mbox{ s}^{-1})$ for each of the seven 12-hour intervals in our study. As shown in Figure~\ref{fig:fig4}, the power spectra calculated from the two methods (top) are consistent with each other. Further, both methods produce consistent local spectral indices (bottom) over the high frequency range, but gradually deviate from each other for $f<2\times 10^{-2} \mbox{ s}^{-1}$. Figure~\ref{fig:fig5} shows that the resulting low-frequency spectral indices decrease from $\simeq -0.5$ to $\simeq -0.9$ as $r$ increases from $17 R_{\rm s}$ to $45 R_{\rm s}$, similar to the spectral indices inferred from the PSDs.

\begin{figure}[ht!]
\centering
\includegraphics[width=6in,height=1.9in]{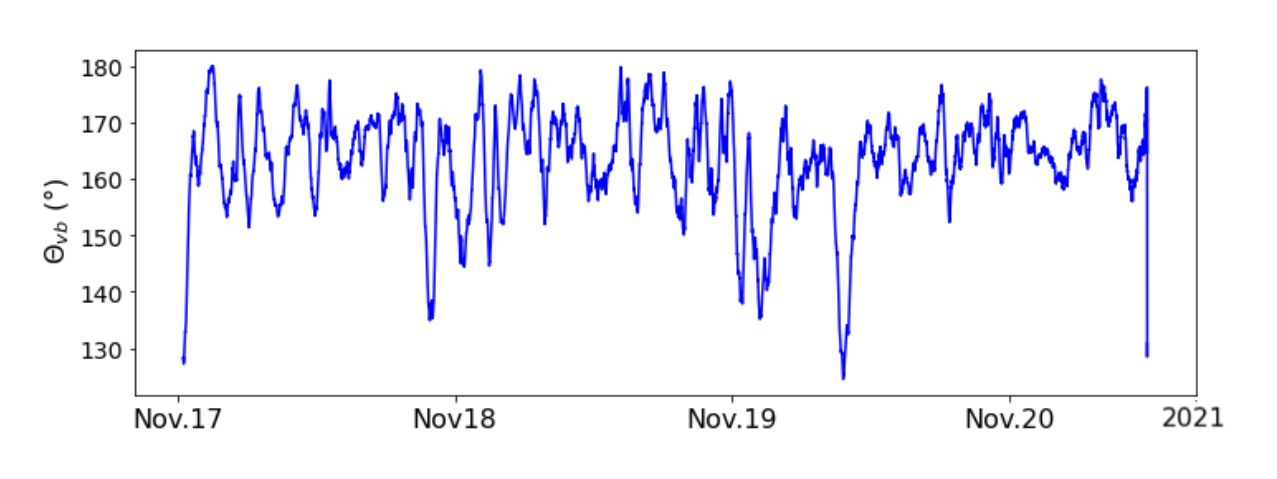}
\caption{ Variation of the angle between the mean magnetic field and mean velocity field in the spacecraft frame ${\theta }_{\mathrm{VB}}^{\mathrm{sc}}$. }
\label{fig:fig6}
\end{figure}

As noted by \cite{kraichnan65}, an Alfv\'en wave packet tends to propagate parallel or anti-parallel to the direction of the local background magnetic field, which is the local spatial average of the magnetic field over a volume several times larger than the Alfv\'en wave packet. For the fluctuations at $f<f_{\rm b}$, this local background magnetic field is comparable to a one-hour average of the magnetic field. According to Taylor's (1938) \nocite{taylor38} hypothesis, temporal variations in the spacecraft frame correspond to spatial variations along the direction of the solar-wind velocity $\bm{V}$ as measured in the spacecraft frame. The frequency spectra that we plot thus correspond to 1D wavenumber spectra along the direction of~$\bm{V}$ -- i.e.,  integrals of the 3D wavenumber spectra over the two wavenumber components orthogonal to~$\bm{V}$. To interpret the frequency spectra, we compute the angle $\theta_{\mathrm{VB}}^{\mathrm{sc}}$
between running one-hour averages of $\bm{B}$ and the spacecraft-frame velocity~$\bm{V}$. As shown in Figure~\ref{fig:fig6},  most of the time  $160 \lesssim {\theta }_{\mathrm{VB}}^{\mathrm{sc}} \lesssim 180 $. Because $\sin(\theta_{\mathrm{VB}}^{\mathrm{sc}})$ is small, the frequency spectra that we measure at $f<f_{\rm b}$ correspond approximately to the {\em parallel}-wavenumber spectra of the low-frequency fluctuations -- i.e., the 3D wavenumber spectra integrated over both of the wavenumber components perpendicular to the local average magnetic field (the direction of which, up to an overall sign, is close to the direction along which fluctuations in the $1/f$ range propagate). The evolution towards a $1/f$ spectrum in this fast-solar-wind stream is thus similar to the evolution towards a $1/k_\parallel$ spectrum that occurs during the nonlinear evolution of the parametric decay instability when slow magnetosonic waves are strongly damped, where $k_\parallel$ is the wavenumber component parallel to the average magnetic field \citep{chandran18a}.


\section{Summary and Conclusion} \label{sec:summary}

In this paper, we have examined magnetic-field data from PSP's encounter 10, specifically from November 17, 2021, to November 20, 2021. During this interval, PSP was moving approximately radially within fast solar wind, as illustrated in Figure~\ref{fig:fig1}, and was magnetically connected to a single mid-latitude coronal hole \citep{badman23}.  
The magnetic power spectra presented here all show an approximate double-power-law form over the frequency range $(10^{-4} \mbox{ s}^{-1}, 10^{0} \mbox{ s}^{-1})$, with a break frequency~$f_{\rm b}$ that gradually decreases as $r$ increases. Our principal finding is that the spectral index at $f<f_{\rm b}$ gradually decreases from $-0.61$  to $-0.94$ as $r$ increases from $17.4 R_{\rm s}$ to $45.7 R_{\rm s}$. Although we have analyzed only a single fast-solar-wind stream, our results suggest that the $1/f$ power spectrum that is observed in fast wind at $f< f_{\rm b}$ at $r< 60 R_{\rm s}$ does not originate at the Sun, but instead evolves dynamically as the solar wind flows from the corona out to $r\sim 60 R_{\rm s}$. Further analysis of more solar-wind streams is needed to determine the extent to which the behavior we find characterizes the fast solar wind as a whole.

If the spectral index at $f< f_{\rm b}$ does indeed evolve dynamically, then determining the physical mechanisms responsible for this evolution is an important unsolved problem. As mentioned in the introduction, previous studies have argued that a $1/f$ scaling can be produced by nonlinear interactions in reflection-driven Alfv\'en-wave turbulence \citep[e.g.][]{velli89,verdini12} or by the nonlinear evolution of the parametric decay instability \citep{chandran18a}. However, further investigation is needed to determine whether these mechanisms, or some other mechanism, can explain the PSP measurements we have presented. This further investigation includes additional work to elucidate the effects of reflection-driven turbulence and parametric decay on the spectrum at $f<f_{\rm b}$ as well as additional observational investigations. In addition to the previously mentioned need to determine whether the radial evolution of $\alpha_{\rm B}$ at $f < f_{\rm b}$ that we have found occurs in a larger sample of fast-solar-wind streams, more work is needed to characterize the radial evolution of other types of fluctuations besides fluctuations of~$\bm{B}$. In particular, it will be important to determine the radial evolution of fluctuations in the  density, velocity, and Elsasser variables~\citep{elsasser50},  which  correspond to Alfv\'en-wave fluctuations propagating towards and away from the Sun in the plasma frame. 




\begin{acknowledgments}
We thank Aaron Roberts for helpful discussions. ND and BC were supported in part by NASA grant NNN06AA01C to the Parker Solar Probe FIELDS Experiment and NASA grant 80NSSC19K0829. CHKC is supported by UKRI Future Leaders Fellowship MR/W007657/1 and STFC Consolidated Grant ST/T00018X/1. We thank the members of the FIELDS/SWEAP teams and PSP community for helpful discussions. PSP data are available at the SPDF (https://spdf.gsfc.nasa.gov).
\end{acknowledgments}

\bibliography{articles.bib}{}
\bibliographystyle{aasjournal}

\end{document}